\documentclass[12pt]{iopart}

\usepackage[margin=1in]{geometry}
\usepackage[titletoc,toc,title]{appendix}
\usepackage{braket}
\usepackage[final]{feynmp}
\usepackage{ifpdf}
\usepackage{comment}
\usepackage{amssymb}
\usepackage{mathrsfs}
\usepackage{color}
\usepackage{bm}
\usepackage{times}

\usepackage{enumerate}
\usepackage{physics}
\usepackage{caption}
\usepackage{subcaption}

\usepackage[demo]{graphicx}% Include figure files
\usepackage{dcolumn}% Align table columns on decimal point
\usepackage{bm}% bold math
\usepackage{hyperref}
\hypersetup{
    colorlinks,
    citecolor=blue,
    filecolor=blue,
    linkcolor=blue,
    urlcolor=blue
}

\makeatletter
\renewcommand*\env@matrix[1][\arraystretch]{%
  \edef\arraystretch{#1}%
  \hskip -\arraycolsep
  \let\@ifnextchar\new@ifnextchar
  \array{*\c@MaxMatrixCols c}}
\makeatother

\newcommand{\beq}{\begin{equation}}
\newcommand{\eeq}{\end{equation}}

\renewcommand{\arraystretch}{1.5}

 \newcommand{\dg}{^\dagger}

\renewcommand{\Re}{\mathrm{Re}}
\renewcommand{\Im}{\mathrm{Im}}

\newcommand{\rar}{\rightarrow}
\newcommand{\dn}{\downarrow}
\newcommand{\up}{\uparrow}

 \def \a {\alpha}

\def\d{\partial}
\def \Tr {\mathrm{Tr}}

\def \be {\begin{equation}}
\def \ee {\end{equation}}
\def \bs {\begin{split}}
\def \es {\end{split}}
\def\bea {\begin{eqnarray}}
\def\eea {\end{eqnarray}}

\def \ha {\hat{a}}

\def \hs {\hat{s}}

\def \Z {\mathcal{Z}}
\def \D {\mathcal{D}}

\begin{document}

\title[]{Nontrivial saddle points for spectral form factors of flat band superconductors}

\author{Sankalp Gaur$^1$, Emil A. Yuzbashyan$^2$ and Victor Gurarie$^1$}

\address{$^1$Department of Physics and Center for Theory of Quantum Matter, University of Colorado, Boulder, Colorado 80309, USA}
\address{$^2$Department of Physics and Astronomy, Center for Materials Theory, Rutgers University, Piscataway, NJ 08854, USA}

\ead{victor.gurarie@colorado.edu}
\vspace{10pt}

\begin{abstract}
We derive the spectral form factor of a flat band superconductor in two different ways. 
In the first approach, we diagonalize the Hamiltonian of this system exactly and numerically sum over the exact eigenstates to find the spectral form factor. In the second approach, we use mean field theory
to evaluate the same spectral form factor. We demonstrate that both methods produce the same answer. Mean field theory for spectral form factors possesses features 
not previously seen in the theory of superconductivity, in particular complex gap functions and non-Hermitian effective Hamiltonians. We explicitly show that these features are indeed necessary to obtain the correct spectral form factor. 
\end{abstract}

%\maketitle

\section{Introduction}
Spectral form factors of quantum systems \cite{MehtaRM}, defined by
\beq   \Z (t) = \Tr \, e^{-i \hat{H} t} = \sum_n e^{-i E_n t} \ ,
\label{eq:Z}
\eeq
where $E_n$ are their energy levels, are closely related to the correlations between the density of states
\beq R(\omega) = 2 \pi \sum_{n,m} \delta \left(\omega-E_n+E_m \right).
\eeq
It is straightforward to see that
\beq R(\omega) = \int_{-\infty}^\infty dt \left| \Z(t) \right|^2 e^{i \omega t}.
\eeq
$\Z(t)$  is obviously a trace of the evolution operator of a quantum system, and can also be thought of as  the analytic continuation of the partition functions to imaginary values of temperature. 

Recently the authors of this manuscript studied the spectral form factor for a variety of superconductors \cite{Gaur2024}. We observed that unconventional superconductors with the gap function which vanishes in part of the Brillouin zone typically have 
nonanalytic spectral form factors at certain times $t$. The singularities in the spectral form factor reflect the underlying structure of the gap function of the superconductor. This is similar to how dynamical quantum phase transitions are related to the underlying equilibrium phases of quantum systems \cite{Kehrein2013, Vajna2015, Heyl2018}.

To obtain these results, we analytically continued the gap function of a superconductor from the values it takes at some temperature $T$ to the values when the temperature is analytically continued as in  $(k_B T)^{-1} \rightarrow i t$.  One striking observation that we had to make was that under the analytic continuation the gap function is no longer real, in the sense that $\bar \Delta \Delta$ becomes a complex number. The nonanalyticity of the spectral form factors crucially relied on this observation. 

It would be advantageous to identify a model of interacting fermions whose ground state resembles that of a superconductor but whose spectral form factor could be independently determined by a direct diagonalization
of its Hamiltonian. We could then compare the result of that calculation with the analytic continuation of the gap function to the complex temperature. This would provide a direct verification 
of the $\bar \Delta \Delta$ acquiring complex values when computed for the purpose of evaluating spectral form factors. 

We identified such a model as being the one given by the following Hamiltonian
\begin{equation} 
    \hat{H} = \epsilon \sum_{n = 1}^N \left( \hat{a}_{n \up}\dg \hat{a}_{n \up} + \hat{a}_{n \dn}\dg\hat{a}_{n \dn}  \right)
    - \frac{g}{N} \sum_{n = 1}^N \sum_{m = 1}^N \hat{a}_{n \up}\dg \hat{a}_{n \dn}\dg \hat{a}_{m \dn} \hat{a}_{m \up} \ 
\label{eq:H}
\end{equation}
with $g>0$. 
$\ha$ and $\ha\dg$ are respectively the annihilation and creation operators for attractively interacting identical fermions. Such a system of fermions has a ground state similar to that of an $s$-wave superconductor \cite{SchreifferSC}. 
Further, in the limit $N \rightarrow \infty$, the BCS-like ground state of this superconductor becomes exact. In what follows, we will study the large $N$ behavior of this Hamiltonian. 

%The index $n$ here represents momentum modes for an $s$-wave superconductor with translational invariance. This is a flat band superconductor since the bare fermion energy $\epsilon$ is independent of $n$. In this article, we work with the above Hamiltonian without worrying about the explicit lattice structure of the underlying superconductor. Hence for the purpose of this calculation, we treat $n$ just as an index and not consider the distribution of momenta in the Brillouin Zone.

We note that an equivalent representation of this Hamiltonian is a collection of interacting two level systems according to
\begin{equation} \label{eq:ham}
    \hat{H} = 2 \epsilon \sum_{n = 1}^N \left( \hs_n^z + \frac{1}{2} \right) 
    - \frac{g}{N} \sum_{k,l = 1}^N \hs_k^+ \hs_l^- \ .
\end{equation}
Here we employ the following standard mapping from fermionic operators to spin operators,
\begin{equation}
    \ha_{n \up}\dg \ha_{n \dn}\dg = \hs_n^+ \quad ,\quad 
    \ha_{n \dn} \ha_{n \up} = \hs_n^- \quad,
    \quad \frac{1}{2} (\ha_{n \up}\dg \ha_{n \up} + \ha_{n \dn}\dg \ha_{n \dn} - 1) = \hs_n^z \ ,
\label{eq:anderpseudo}
\end{equation}
where
\begin{equation}
    \hs_n^+ = \hs_n^x + i \hs_n^y \quad , \quad
    \hs_n^- = \hs_n^x - i \hs_n^y \ .
\end{equation}
These operators follow the commutation relations, 
\begin{equation}
[\hs_m^\mu, \hs_n^\nu] = i \delta_{mn} \sum_{\rho} \epsilon^{\mu \nu \rho} \hs_m^{\rho}
\end{equation}
(where the Greek indices stand for $x$, $y$ and $z$)
and are called Anderson pseudospins \cite{Anderson1958}.
In this form, this Hamiltonian can be realized in cavity QED experiments \cite{Young2024}. 

If we are to study the Hamiltonian (\ref{eq:ham}) at a finite temperature, we can derive the standard gap equation for the superconducting gap function $\Delta$ at a temperature $T$ \cite{SchreifferSC, Leggett1975},
\beq \label{eq:realgap} \frac{g}{2} \frac{\tanh \left[ \frac{\sqrt{\epsilon^2+ \bar \Delta \Delta}  }{k_B T} \right] }{\sqrt{\epsilon^2 + \bar \Delta \Delta}}= 1.
\eeq
Famously, its solutions are nonzero $\bar \Delta \Delta>0$ at a low enough temperature $T$ (if $g/(2 \epsilon)>1$), and become zero $\bar \Delta \Delta=0$ at $T=T_c$, the critical temperature of the onset of superconductivity. The superconducting gap function is formally defined below in ~(\ref{eq:action}). In the context of cavity QED realization of this model, this transition is sometimes referred to as the superradiance transition.

For the purpose of evaluating spectral form factors we need to analytically continue this equation to imaginary $T$, or $(k_B T)^{-1} \rightarrow it$. It should be clear that no real nonzero $\bar \Delta \Delta$ 
can satisfy the gap equation (\ref{eq:realgap}) if $T$ is imaginary. Therefore, we expect to see solutions to (\ref{eq:realgap}) where $\bar \Delta \Delta$ is complex. 

In this paper we compute the spectral form factor of the Hamiltonian (\ref{eq:ham}) in two ways: by a direct diagonalization followed by a numerical summation over the states in ~(\ref{eq:Z}), and by 
computing the gap function employing the gap equation. We show that both give the same answer. We show that the gap function needed for evaluating the spectral form factor is indeed complex, that is, $\bar \Delta \Delta$ takes complex values. 

\section{Mean field approximation}

The standard way to work with a superconducting Hamiltonian is to introduce a mean field. The equilibrium mean field corresponds to the superconducting gap function. Here we shall set up the mean field to calculate the spectral form factor. From (\ref{eq:Z}) and (\ref{eq:H}), we can write $\Z(t)$ as a coherent state path integral \cite{Coleman_Many_Body}, 
\begin{equation}
        \Z (t) = \int \D \psi \D \bar{\psi}\, \exp \left[ i \int_0^t d \tau \left( \sum_{n , \sigma} \left( i \bar{\psi}_{n \sigma} \dot{\psi}_{n \sigma} - \epsilon \bar{\psi}_{n \sigma} \psi_{n \sigma} \right) + \frac{g}{N} \sum_{n, m} \bar{\psi}_{n \up} \bar{\psi}_{n \dn} \psi_{m \dn} \psi_{m \up} \right) \right] \ ,
\label{eq:path}
\end{equation}
where $\psi_{n \sigma}$ and $\bar{\psi}_{n \sigma}$ are fermionic fields. In order for this to represent the trace, these fields must satisfy the boundary conditions,
\begin{equation}
    \psi_{n \sigma} (t) = - \psi_{n \sigma} (0) \quad ,\quad 
    \bar{\psi}_{n \sigma} (t) = -\bar{\psi}_{n \sigma} (0) \ .
\end{equation}
%We have neglected subexponential prefactors in \eqref{eq:path} as we are only trying to find the mean field here. 
Applying the Hubbard-Stratonovich transformation \cite{Coleman_Many_Body, Hubbard1959} \eqref{eq:path} becomes
\begin{equation}
\begin{split}
    \Z (t) &= \int \D \psi \D \bar{\psi} \D \Delta \D \bar{\Delta} \, e^{i S(t)} \quad , \\
    S(t) &= \int_0^t d \tau \left( \sum_{n, \sigma} \left( i \bar{\psi}_{n \sigma} \dot{\psi}_{n \sigma} - \epsilon \bar{\psi}_{n \sigma} \psi_{n \sigma} \right) + \Delta \sum_n \bar{\psi}_{n \up} \bar{\psi}_{n \dn} + \bar{\Delta} \sum_n \psi_{n \dn} \psi_{n \up} - \frac{N}{g} \bar{\Delta} \Delta \right) \ .
\end{split}
\label{eq:action}
\end{equation}
We can rewrite this functional integral by defining the effective action $W$ for a single species of fermions $\psi$ via
\beq e^{i W } =  \int \D \psi \D \bar{\psi} \exp \left( i \int_0^t dt  \left[ \sum_\sigma \left( i \bar{\psi}_{ \sigma} \dot{\psi}_{ \sigma} - \epsilon \bar{\psi}_{ \sigma} \psi_{ \sigma} \right) + \Delta  \bar{\psi}_{n \up} \bar{\psi}_{n \dn} + \bar{\Delta} \psi_{\dn} \psi_{ \up} \right] \right),
\eeq
with the result
\beq \Z(t) = \int \D \Delta \D \bar \Delta \, e^{i N \left(W - \frac{\bar \Delta \Delta}{g} \right) }.
\eeq
This form makes it clear that at large $N$ mean field approximation, which consists of computing the extremum of $W - \bar \Delta \Delta/g$ over $\Delta$ and $\bar \Delta$, becomes exact. 
We can also refer to this procedure as saddle point approximation. 
Therefore we approximate the functional integral over $\Delta(\tau)$ and $\bar\Delta(\tau)$ using only their saddle points. This calculation of the saddle points is detailed in \ref{sec:saddle}.

We see that there are two classes of time-dependent saddle point solutions, given by \eqref{eq:sp1} and \eqref{eq:sp2}. The former solution has $\bar \Delta = \Delta=0$, while the latter
have nonzero $\bar \Delta$ and $\Delta$, with $\bar \Delta \Delta$ being complex. We calculate the contribution to the spectral form factor due to each. Substituting \eqref{eq:sp1} in \eqref{eq:z0f} gives
\begin{equation}
    \Z_0 (t) = 2 \cos (\epsilon t)
\label{eq:spd1}
\end{equation}
whereas substituting the solution \eqref{eq:sp2} gives
\begin{equation}
    \Z_0 (t) = 2 (-1)^k \cos (t z_l (t)) \exp \left[ - \frac{i t}{g}  \left( z_l (t)^2 - \left( \frac{k \pi}{t} - \epsilon \right)^2 \right) \right]\ ,
\label{eq:spd2}
\end{equation}
where \begin{equation} \Z_0 (t) = (\Z(t))^{\frac{1}{N}}, \end{equation} as defined in \eqref{eq:z0n} and \eqref{eq:z0}, is the spectral form factor for a single fermionic mode of the mean field Hamiltonian, and $z_l (t)$ is a solution for $z$ in the equation
\begin{equation}
    \frac{i}{2} \frac{\tan (z t)}{z} = \frac{1}{g} \ .
\label{eq:spd0}
\end{equation}
A parameter $k$ which takes integer values is arbitrary (it corresponds to choosing different time dependence of the saddle point solutions, see ~(\ref{eq:timedep})). The absolute value of ${\Z_0}$ does not depend on it, so it can remain arbitrary in our approach. 
\eqref{eq:spd0} has infinitely many solutions in the complex plane, which we index with subscript $l$. Note that these saddle points $z_l(t)$ depend on the parameters $g$ and $t$, but not on $\epsilon$.
\eqref{eq:spd0} is transcendental, so we cannot obtain these solutions analytically. Nonetheless, we can evaluate them numerically and describe their dependence on the parameters $g$ and $t$ qualitatively. 

\begin{figure}[h]
    \centering
    \includegraphics[width=0.9\linewidth]{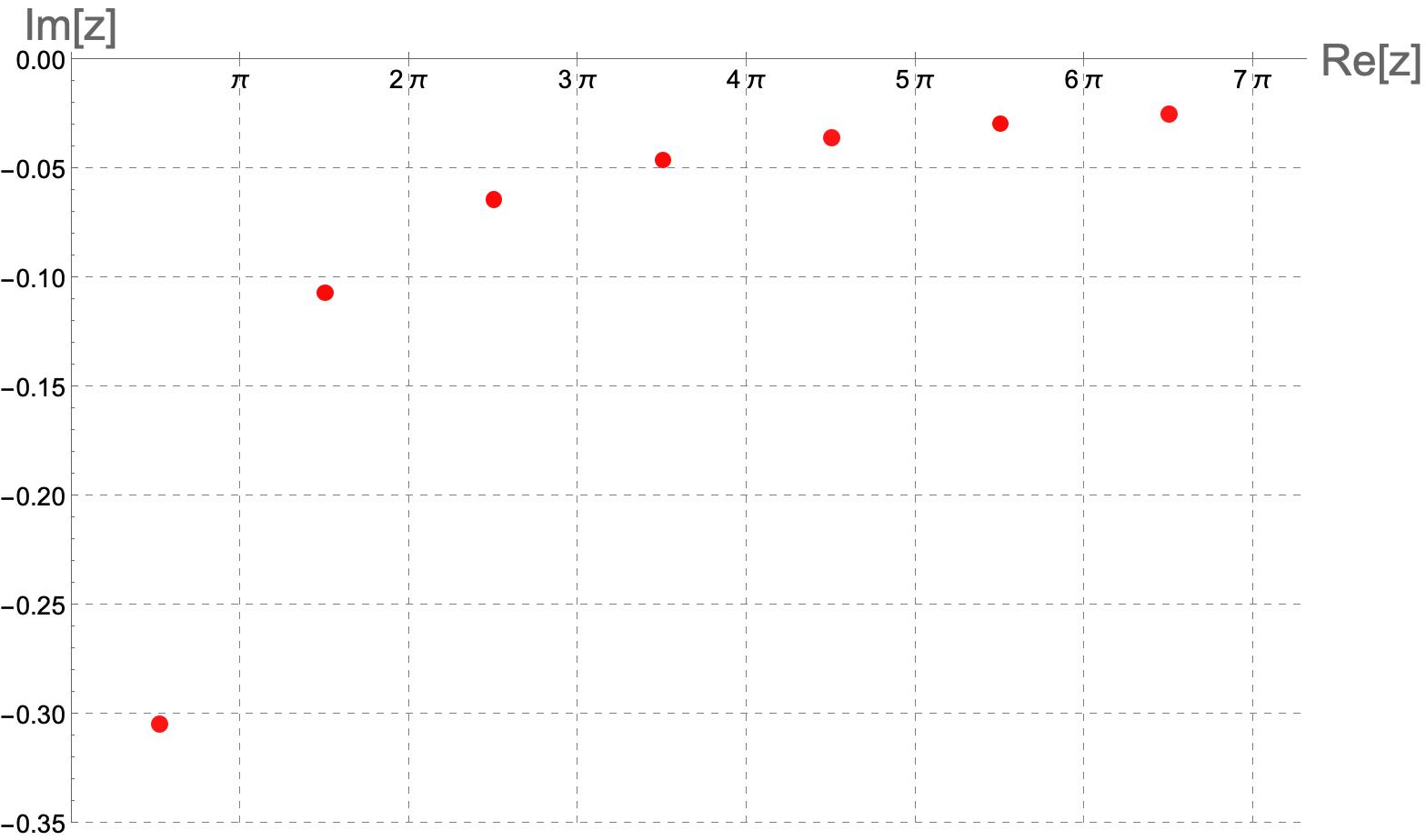}
    \caption{Saddle points $z_l(t)$ for g = 1 and t = 1. Shown here are the seven saddle points closest to the imaginary axis in the lower right quadrant of the complex plane.}
    \label{fig:saddle_t_1}
\end{figure}

Figure \ref{fig:saddle_t_1} shows the saddle points $z_l(t)$ which are the solutions to \eqref{eq:spd0} for $g = 1$ and $t = 1$. The seven saddle points closest to the imaginary axis in the lower right quadrant of the complex plane are plotted. For each saddle point $z_l(t)$, there is a corresponding saddle point $-z_l(t)$ in the upper left quadrant of the complex plane not shown in the figure.

\begin{figure}
    \centering
    \includegraphics[width=1\linewidth]{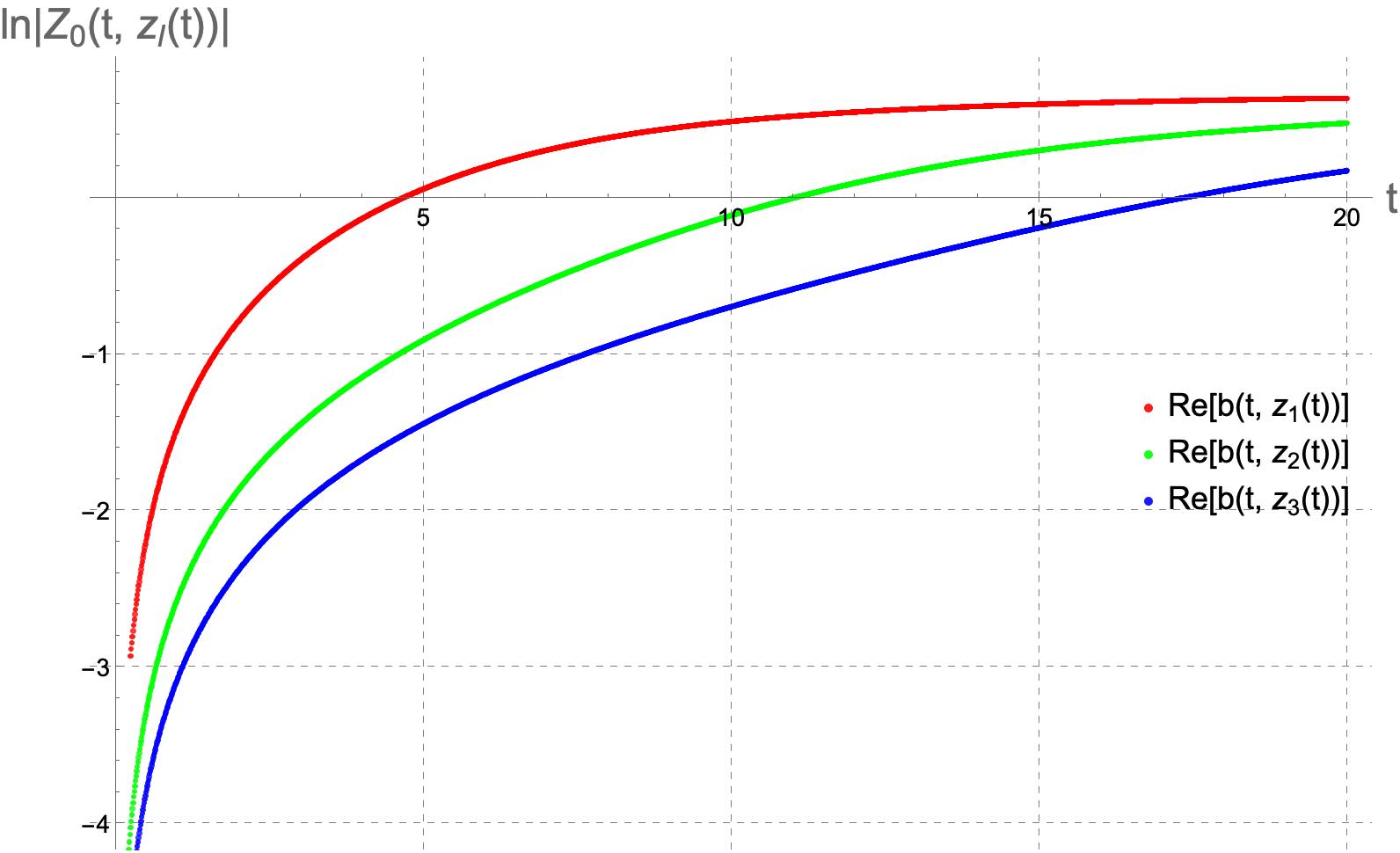}
    \caption{Contribution of saddle points $z_1$, $z_2$ and $z_3$ to $\Z_0$ as a function of time $t$ for $g = 1$}
    \label{fig:saddle3}
\end{figure}

The infinitely many saddle point solutions to \eqref{eq:spd0}, each contribute to the spectral form factor according to \eqref{eq:spd2}. Let us define 
\begin{equation}
    b(t, z_l(t)) = \ln (\Z_0(t, z_l(t))) \ ,
\label{eq:b}
\end{equation}
where the expression for $\Z_0 (t, z_l (t))$ is given in \eqref{eq:spd2}. The magnitude of the contribution of saddle point $z_l(t)$ to the spectral form factor is determined by the real part of $b(t, z_l(t))$,
\begin{equation}
    \Re[b(t, z_l(t))] = \ln 2 + \ln |\cos(t z_l(t))| + \frac{2 t}{g} \Re[z_l(t)] \Im[z_l(t)] \ .
\label{eq:reb}
\end{equation}
whereas the imaginary part determines its complex phase. Denoting the saddle points in the lower right quadrant of the complex plane as $z_1(t),\ z_2(t),\ z_3(t)$, and so on according to their distance from the imaginary axis, the real part of $b(t, z_l(t))$ is shown as a function of $t$ for the first three saddle points in Figure \ref{fig:saddle3}. The interaction strength $g$ has been set to $1$ for this figure. We see that $z_1(t)$, the saddle point closest to the imaginary axis, has the largest value of $\Re[b(t, z_l(t)]$ among all the saddle points.

Summing over all the saddle points the spectral form factor is obtained from \eqref{eq:z0n} and \eqref{eq:b} as 
\begin{equation}
    \Z(t) = \sum_l c_l (t) \, e^{N b(t, z_l(t))} \ ,
\label{eq:saddle_sum}
\end{equation}
where $c_l(t)$ is a prefactor that is subexponential in $N$. $c_l(t)$ comes from the subexponential prefactor in \eqref{eq:path} and the Gaussian integral around the saddle point. Thus from \eqref{eq:saddle_sum} and Figure \ref{fig:saddle3} we see that saddle points $z_l (t)$ with $l > 1$ are exponentially suppressed as compared to the saddle point $z_1(t)$. To get the spectral form factor from \eqref{eq:spd2} in the large-$N$ limit, we only consider $z_l(t)$ with $l=1$ at the leading order.

Among \eqref{eq:spd1} and \eqref{eq:spd2}, at leading order in $N$, we only consider the saddle point whose contribution to $\Z_0(t)$ has the larger absolute value. Using \eqref{eq:spd1}, \eqref{eq:spd2}, \eqref{eq:b}, and \eqref{eq:reb}, we obtain
\begin{equation}
    \lim_{N \rar \infty} \frac{\ln |\Z(t)|}{N} = \ln 2 + \max \left\{ \ln (|\cos(\epsilon t)|), \left( \ln |\cos(t z_1(t))| + \frac{2 t}{g} \Re[z_1(t)] \Im[z_1(t)] \right) \right\} \ .
\label{eq:saddle_final}
\end{equation}
This is an exact expression for the spectral form factor of the Hamiltonian in \eqref{eq:H}. The term $\ln \left( \left| \cos \left(\epsilon t \right) \right| \right)$ represents the contribution of the 
trivial saddle point $\bar \Delta= \Delta =0$. The second  term represents the contribution of the nontrivial saddle point $z_1(t)$. 

\section{Exact solution: sum over Anderson pseudospins}

%There exists the following standard mapping from fermionic operators to spin operators,
%\begin{equation}
 %   \ha_{n \up}\dg \ha_{n \dn}\dg = \hs_n^+ \quad ,\quad 
  %  \ha_{n \dn} \ha_{n \up} = \hs_n^- \quad,
   % \quad \frac{1}{2} (\ha_{n \up}\dg \ha_{n \up} + \ha_{n \dn}\dg \ha_{n \dn} - 1) = \hs_n^z \ ,
%\end{equation}
%where
%\begin{equation}
  %  \hs_n^+ = \hs_n^x + i \hs_n^y \quad , \quad
    %\hs_n^- = \hs_n^x - i \hs_n^y \ .
%\end{equation}
%These operators follow the commutation relations, 
%\begin{equation}
 %   [\hs_n^x, \hs_n^y] = i \hs_n^z \quad, \quad
   % [\hs_n^y, \hs_n^z] = i \hs_n^x \quad, \quad
   % [\hs_n^z, \hs_n^x] = i \hs_n^y\ ,
%\end{equation}
%and are called Anderson pseudospins [cite].
Let us now examine our problem as written in terms of Anderson pseudospins, as in ~\eqref{eq:ham}. 
%In terms of Anderson pseudospins, the Hamiltonian \eqref{eq:H} becomes
%\begin{equation}
 %   \hat{H} = 2 \epsilon \sum_{n = 1}^N \left( \hs_n^z + \frac{1}{2} \right) 
   % - \frac{g}{N} \sum_{k,l = 1}^N \hs_k^+ \hs_l^- \ .
%\end{equation}
Denoting the sum over all pseudospins as $\hat{\mathbf{S}} = \sum_{n=1}^N \hat{\mathbf{s}}_n$, the Hamiltonian becomes
\begin{equation}
    \hat{H} = N \epsilon + 2 \epsilon \hat{S}^z - \frac{g}{N} \hat{S}^+ \hat{S}^- \ .
\label{eq:HSpin}
\end{equation}
Since $\hat{S}^+ \hat{S}^- = \hat{\mathbf{S}}^2 - (\hat{S}^z)^2 + \hat{S}^z$ and $[\hat{\mathbf{S}}^2, \hat{S}_z] = 0$, eigenstates of the Hamiltonian \eqref{eq:HSpin} are simultaneous eigenstates of $\hat{\mathbf{S}}^2$ and $\hat{S}_z$ operators. Let us denote these eigenstates as $\ket{K M}$, where
\begin{equation}
    \hat{\mathbf{S}}^2 \ket{K M} = K (K+1) \ket{K M} \quad , \quad 
    \hat{S}^z \ket{K M} = M \ket{K M} \ .
\end{equation}
If we choose $N$ even for convenience, then the value of $K$ can be any integer from $0$ to $N/2$. For a particular $K$, $M$ can have integer values from $-K$ to $K$. The possible eigenvalues of the Hamiltonian \eqref{eq:HSpin} are hence,
\begin{equation}
    E_{K M} = N \epsilon + 2 \epsilon M - \frac{g}{N} [K (K+1) - M(M-1)] \ ,
    \label{eq:h1}
\end{equation}
each with degeneracy \cite{Landau9},
\begin{equation}
    D_K = \frac{N! (2K + 1)}{\left(\frac{N}{2} + K + 1\right)! \left(\frac{N}{2} - K \right)!} \ .
    \label{eq:h2}
\end{equation}
This gives us an expression for the spectral form factor \eqref{eq:Z} as 
\begin{equation}
    \Z(t) = e^{- i N \epsilon t} \sum_{K = 0}^{\frac{N}{2}} \frac{N! (2K + 1)}{\left(\frac{N}{2} + K + 1\right)! \left(\frac{N}{2} - K \right)!}
    e^{i t \frac{g}{N} K(K+1)}
    \sum_{M = - K}^K
    e^{-i t \left( 2 \epsilon M + \frac{g}{N} M (M-1) \right)} \ .
\label{eq:sumoverspins}
\end{equation}

\begin{figure}[h!]
\captionsetup[subfigure]{justification=centering}
    \centering
    \begin{subfigure}{0.49\linewidth}
        \includegraphics[width=\linewidth]{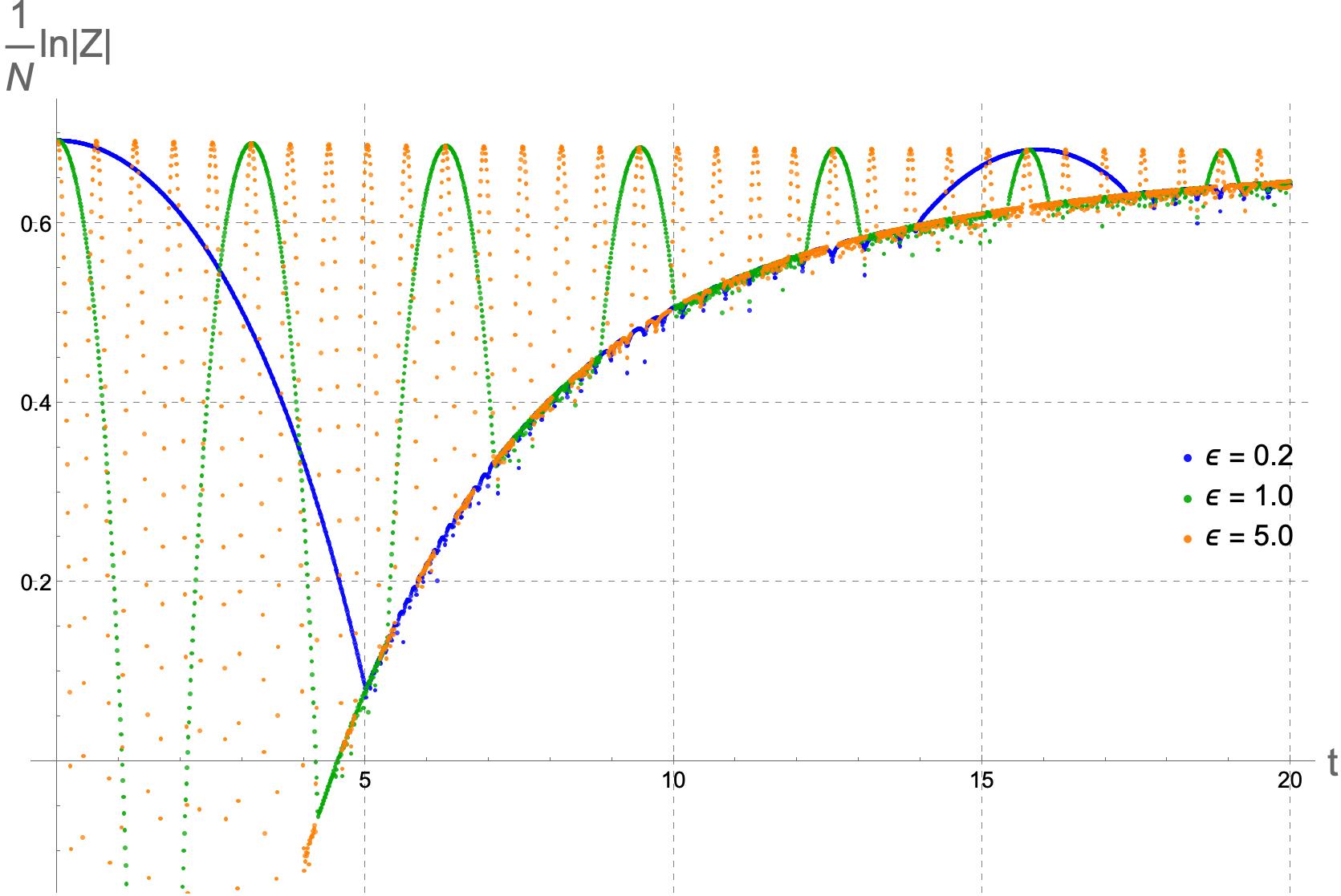}
        \caption{}
        \label{subfig:exact3}
    \end{subfigure}
    \hfill
    \begin{subfigure}{0.49\linewidth}
        \includegraphics[width=\linewidth]{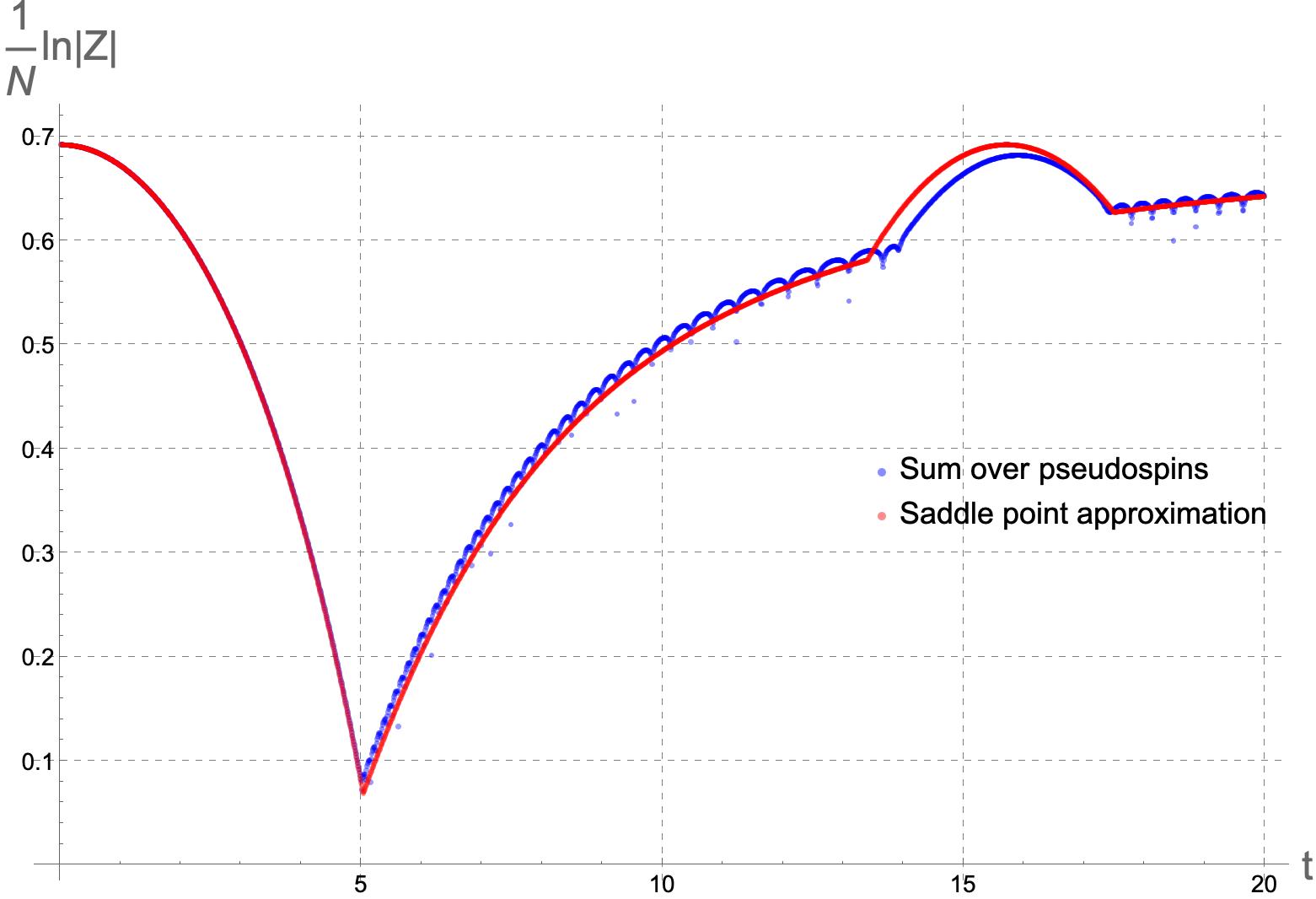}
        \caption{}
        \label{subfig:sffm}
    \end{subfigure}

    \vfill

    \begin{subfigure}{0.49\linewidth}
        \includegraphics[width=\linewidth]{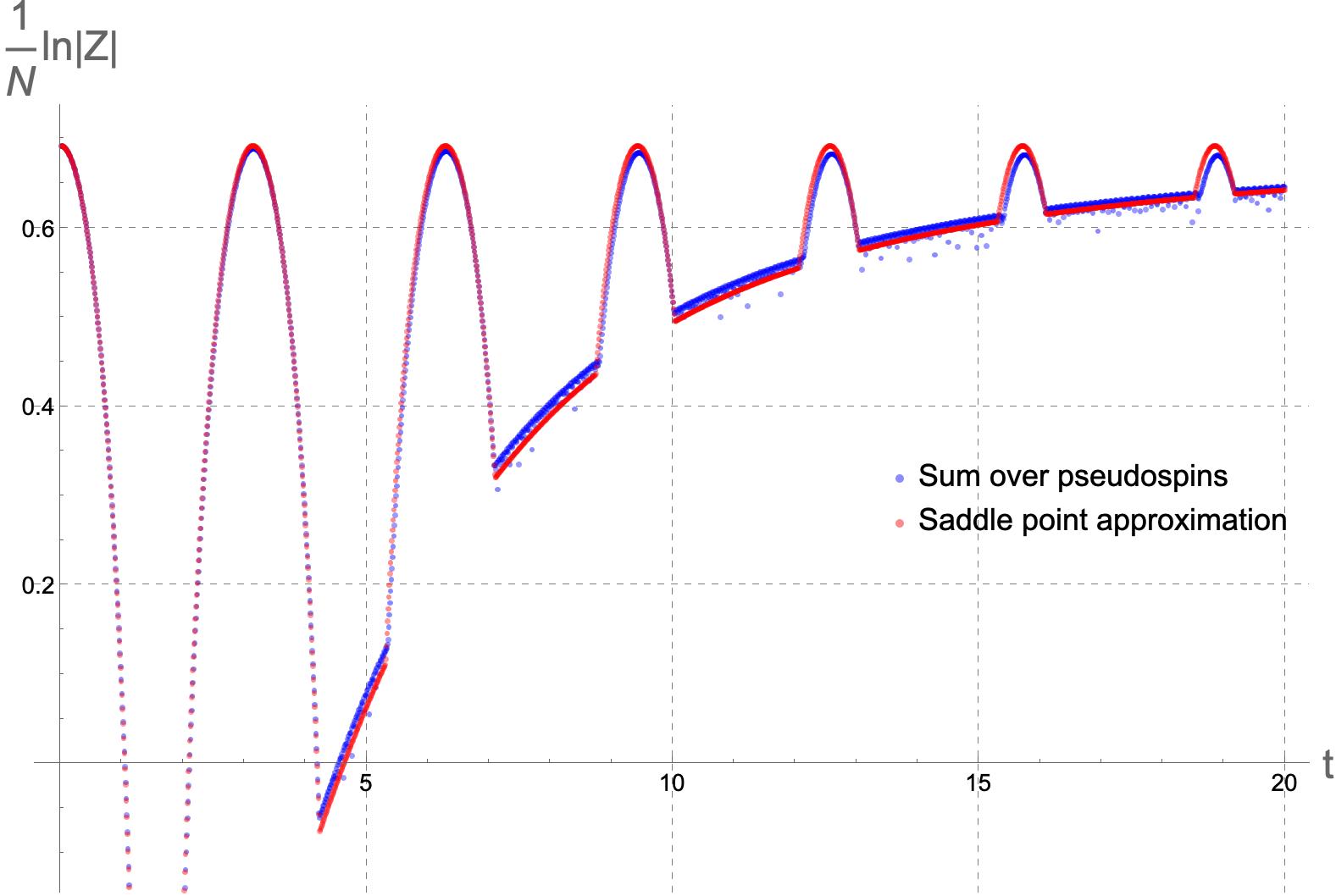}
        \caption{}
        \label{subfig:sff1}
    \end{subfigure}
    \hfill
    \begin{subfigure}{0.49\linewidth}
        \includegraphics[width=\linewidth]{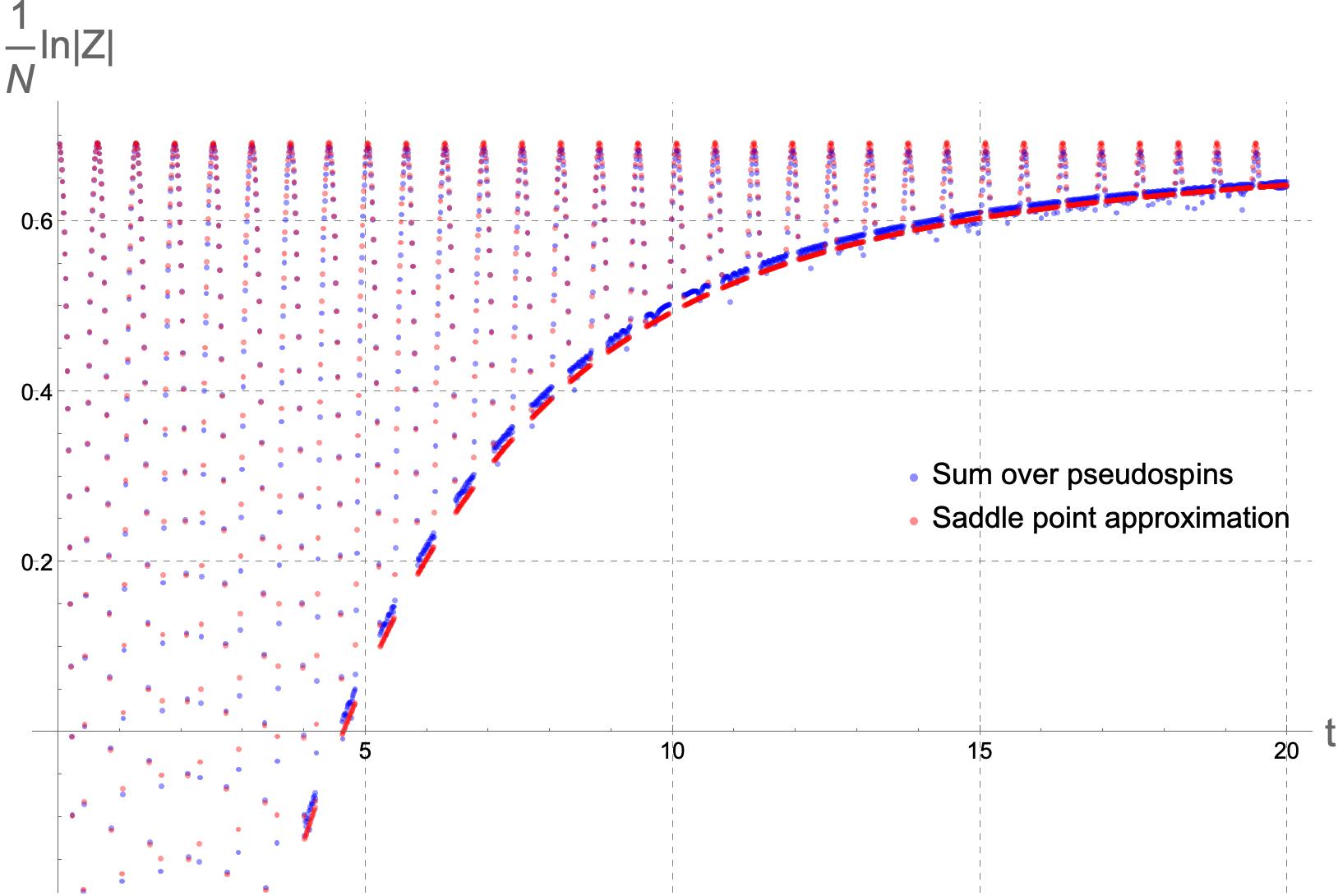}
        \caption{}
        \label{subfig:sffp}
    \end{subfigure}
    
    \caption{(\protect\subref{subfig:exact3}) Exact spectral form factor for three different values of $\epsilon$ with $g=1$, $N=200$. (\protect\subref{subfig:sffm}-\protect\subref{subfig:sffp}) Comparison of the exact spectral form factor from \eqref{eq:sumoverspins} with the saddle point approximation \eqref{eq:saddle_final} with $g =1$, $N = 200$, and (\protect\subref{subfig:sffm}) $\epsilon = 0.2$, (\protect\subref{subfig:sff1}) $\epsilon = 1$, (\protect\subref{subfig:sffp}) $\epsilon = 5$.}
    
    \label{fig:exact_vs_saddle}
\end{figure}

Figure \ref{fig:exact_vs_saddle}(\protect\subref{subfig:exact3}) shows the plot of the absolute value of the spectral form factor from \eqref{eq:sumoverspins} with $g=1$ and $N=200$, for three different values of the bare fermion energy $\epsilon$. Figures \ref{fig:exact_vs_saddle}(\protect\subref{subfig:sffm} - \protect\subref{subfig:sffp}) show the plots of the absolute values of the exact spectral form factor \eqref{eq:sumoverspins} and the saddle point approximation \eqref{eq:saddle_final} with $g=1$ and $N=200$, for $\epsilon = 0.2$, $\epsilon = 1$ and $\epsilon = 5$ respectively. From the plots we see that the saddle point approximation matches the exact spectral form factor to leading order in $N$. The subleading terms are $\mathcal{O}\left(\frac{\ln N}{N}\right)$. For $N=200$, we have $\frac{\ln N}{N} \approx 0.03$.

For saddle point $z_1(t)$, the Hubbard-Stratonovich mean fields satisfy, from \eqref{eq:sp2}, $\bar{\Delta} \Delta = z_1(t)^2 - \left(\frac{k \pi}{t} - \epsilon \right)^2$ for some integer $k$. $z_1(t)$ is neither purely real nor purely imaginary for any $t>0$ since it is a solution for $z$ in \eqref{eq:spd0}. Hence we get an important observation here that $\bar{\Delta} \Delta$ is complex-valued and not purely real. This is in contrast to the mean-field approximation for superconductors in thermal equilibrium, where $\bar{\Delta} \Delta$ is purely real \cite{Coleman_Many_Body}. When calculating the spectral form factor, the required mean-field $\Delta$ does not correspond to the equilibrium gap function of the superconductor. This agrees with our previous results in \cite{Gaur2024}.

Quite interestingly, the spectral form factor as a function of time $t$ exhibits two types of behaviors. One is given by $\ln \left( \left| \cos\left(\epsilon t \right) \right| \right)$ and can be seen as  periodic
upside down parabola-like curves. The other, given by the nontrivial saddle point, is the smooth curve cutting off the contribution of the trivial saddle points from below. 
The derivative of the spectral form factor over time $t$ shows periodic discontinuities as one type of the behavior gets replaced by the other type. 

\section{Discussion}
We have analyzed the spectral form factor of a system of  attractively interacting fermions, with the Hamiltonian (\ref{eq:H}). We have shown that its spectral form factor is a continuous function of time with
periodic discontinuities in its derivatives, which is a typical behavior of spectral form factors in many non-chaotic systems. More importantly, we illustrated that the gap function, defined for the purpose of evaluating the
spectral form factors, become complex, in the sense of $\bar \Delta \Delta$ becoming a complex number. This is also the case when evaluating the Loschmidt echo of a superconductor, which is a dynamical quantity similar to the spectral form factor \cite{Gaur2022}.

The result here comes from the important fact that though the Hubbard-Stratonovich fields defined in ~\eqref{eq:action} are complex conjugates of each other, their saddle points are not. This is also what is observed when a complex function is integrated over a segment of the real axis. If the integrand is complex-valued, then it can have saddle points anywhere in the complex plane, and not necessarily on the real axis \cite{BenderOrszag}.

One conceptual question that this calculation raises is the nature of the Green's functions which arise in the calculation of the gap functions. Those are defined, by analogy with (\ref{eq:z1bar}), as 
\begin{equation}
   G_{\sigma_1 \sigma_2}(\tau_1, \tau_2; t) = \frac{1}{\Z_0(t)} \sum_{\{ \alpha_n \}} \left< {\alpha_n} \right| U (t, \tau_1)\, \ha_{ n \sigma_1}\dg  \, U(\tau_1, \tau_2) \, \ha_{ n \sigma_2 }  \,
   U(\tau_2,0) \left| {\alpha_n} \right>.
\label{eq:z1bar2}
\end{equation}
where 
\be U(\tau_2, \tau_1) = T \exp \left( - i \int_{\tau_1}^{\tau_2} d \tau' \hat{H}_n (\tau') \right),
\ee
while $\hat{ H}_n(\tau)$ and $\Z_0(t)$ are in turn defined in (\ref{eq:hntau}) and (\ref{eq:z0}) respectively. Here $n$ can be chosen to take any particular value from 1 to $N$ and the resulting expression is $n$-independent.  

It follows from the arguments presented here that this Green's function, in the frequency domain, has poles when $\omega \rightarrow \sqrt{\epsilon^2 + \bar \Delta \Delta}$. These poles occur at complex values of 
$\omega$. Usually, the poles of the Green's functions can be interpreted as excitation energies, and they always take real values \cite{Coleman_Many_Body}. In the formalism employed here $\omega \rightarrow \sqrt{\epsilon^2 + \bar \Delta \Delta}$ is complex. The meaning of complex poles of the Green's functions deserves further study. 

Recently a protocol has been developed to measure the spectral form factor of a system of quantum spins \cite{Joshi2022}. The spectral form factor of a superconducting Hamiltonian can be measured with this protocol if the Anderson pseudospins \eqref{eq:anderpseudo} are realized using a trapped ion framework. If we wish to experimentally verify the mean-field result presented here, we need to recall that the expression in ~\eqref{eq:saddle_final} is only valid in the large-$N$ limit. Quantum simulators which measure the spectral form factor are limited by the quantum measurement technology of the time. Currently, the spectral form factor can be measured accurately only for systems with the number of spins $N$ on the order of $10$ \cite{Joshi2022}. In order to compare the theoretical prediction with experiment, ~\eqref{eq:saddle_final} must be extended beyond the large-$N$ limit. This involves including the sub-dominant saddle points in the calculation, the subexponential prefactor to the path integral in ~\eqref{eq:action}, and doing a Taylor expansion of the integrand around each saddle point to include the higher order terms.

\ack
This work was partially supported by
 the Simons Collaboration on Ultra-Quantum Matter,
which is a grant from the Simons Foundation (651440).

\section*{References}
\bibliography{bibliography.bib}

\providecommand{\newblock}{}
\begin{thebibliography}{10}
\expandafter\ifx\csname url\endcsname\relax
  \def\url#1{{\tt #1}}\fi
\expandafter\ifx\csname urlprefix\endcsname\relax\def\urlprefix{URL }\fi
% \eprint[archive=arXiv]{identifier}
\providecommand{\eprint}[2][arXiv]{#1:\linebreak[0]#2}
% Bibliography created with iopart-num v2.1+

\bibitem{MehtaRM}
Mehta M~L 1967 {\em Random Matrices\/} (Elsevier Science \& Technology Books)

\bibitem{Gaur2024}
Gaur S and Gurarie V 2024 Spectral form factors of unconventional
  superconductors {\em Phys. Rev.\/} B {\bf
  \href{https://doi.org/10.1103/PhysRevB.109.144514}{109}}
  \href{https://doi.org/10.1103/PhysRevB.109.144514}{144514}

\bibitem{Kehrein2013}
Heyl M, Polkovnikov A and Kehrein S 2013 Dynamical quantum phase transitions in
  the transverse-field {Ising} model {\em Phys. Rev. Lett.\/} {\bf
  \href{https://doi.org/10.1103/PhysRevLett.110.135704}{110}}
  \href{https://doi.org/10.1103/PhysRevLett.110.135704}{135704}

\bibitem{Vajna2015}
Vajna S and D\'ora B 2015 Topological classification of dynamical phase
  transitions {\em Phys. Rev. B\/} {\bf
  \href{https://doi.org/10.1103/PhysRevB.91.155127}{91}}
  \href{https://doi.org/10.1103/PhysRevB.91.155127}{155127}

\bibitem{Heyl2018}
Heyl M 2018 Dynamical quantum phase transitions: a review {\em Reports on
  Progress in Physics\/} {\bf
  \href{https://dx.doi.org/10.1088/1361-6633/aaaf9a}{81}}
  \href{https://dx.doi.org/10.1088/1361--6633/aaaf9a}{054001}

\bibitem{SchreifferSC}
Schrieffer J~R 1999 {\em Theory of superconductivity\/} (CRC press)

\bibitem{Anderson1958}
Anderson P~W 1958 Random-phase approximation in the theory of superconductivity
  {\em Phys. Rev.\/} {\bf \href{https://doi.org/10.1103/PhysRev.112.1900}{112}}
  \href{https://doi.org/10.1103/PhysRev.112.1900}{1900--1916}

\bibitem{Young2024}
Young D~J, Chu A, Song E~Y, Barberena D, Wellnitz D, Niu Z, Sch{\"a}fer V~M,
  Lewis-Swan R~J, Rey A~M and Thompson J~K 2024 Observing dynamical phases of
  {BCS} superconductors in a cavity {QED} simulator {\em Nature\/} {\bf
  \href{https://rdcu.be/efe8R}{625}} \href{https://rdcu.be/efe8R}{679--684}

\bibitem{Leggett1975}
Leggett A~J 1975 A theoretical description of the new phases of liquid
  $^{3}\mathrm{He}$ {\em Rev. Mod. Phys.\/} {\bf
  \href{https://doi.org/10.1103/RevModPhys.47.331}{47}}
  \href{https://doi.org/10.1103/RevModPhys.47.331}{331--414}

\bibitem{Coleman_Many_Body}
Coleman P 2016 {\em Introduction to Many-Body Physics\/} (Cambridge University
  Press)

\bibitem{Hubbard1959}
Hubbard J 1959 Calculation of partition functions {\em Phys. Rev. Lett.\/} {\bf
  \href{https://doi.org/10.1103/PhysRevLett.3.77}{3}}
  \href{https://doi.org/10.1103/PhysRevLett.3.77}{77--78}

\bibitem{Landau9}
Landau L and Lifshitz E 1977 Chapter {IX} - {Identity} of particles {\em
  Quantum Mechanics (Third Edition)\/} (Pergamon) pp
  \href{https://www.sciencedirect.com/science/article/pii/B9780080209401500165}{225--248}

\bibitem{Gaur2022}
Gaur S, Gurarie V and Yuzbashyan E~A 2022 Singularities in the {Loschmidt} echo
  of quenched topological superconductors {\em Phys. Rev. B\/} {\bf
  \href{https://doi.org/10.1103/PhysRevB.106.L220506}{106}}
  \href{https://doi.org/10.1103/PhysRevB.106.L220506}{L220506}

\bibitem{BenderOrszag}
Bender C~M and Orszag S~A 1999 Asymptotic expansion of integrals {\em Advanced
  Mathematical Methods for Scientists and Engineers I: Asymptotic Methods and
  Perturbation Theory\/} (Springer New York) pp
  \href{https://rdcu.be/ee45n}{247--316}

\bibitem{Joshi2022}
Joshi L~K, Elben A, Vikram A, Vermersch B, Galitski V and Zoller P 2022 Probing
  many-body quantum chaos with quantum simulators {\em Phys. Rev. X\/} {\bf
  \href{https://doi.org/10.1103/PhysRevX.12.011018}{12}}
  \href{https://doi.org/10.1103/PhysRevX.12.011018}{011018}

\end{thebibliography}

\appendix
\section{Saddle point equations}
\label{sec:saddle}

The Hubbard-Stratonovich (HS) mean field Hamiltonian can be written as
\begin{equation}
\begin{split}
    \hat{H} (\tau) &= \sum_{n = 1}^N \hat{H}_n (\tau) \ ,\\
    \hat{H}_n (\tau) &= \epsilon \left(\ha_{n \up}\dg \ha_{n \up} + \ha_{n \dn}\dg \ha_{n \dn} \right) - \Delta(\tau)  \ha_{n \up}\dg \ha_{n \dn}\dg - \bar{\Delta}(\tau)  \ha_{n \dn} \ha_{n \up} + \frac{1}{g} \bar{\Delta}(\tau) \Delta(\tau) \ ,
\end{split}
\label{eq:hntau}
\end{equation}
where $\Delta(\tau)$ and $\bar\Delta(\tau)$ are the HS mean fields at time $\tau$. This is a time($\tau$)-dependent Hamiltonian quadratic in the fermion operators where the different $n-$modes are effectively decoupled. The spectral form factor can be written as a product,
\begin{equation}
\Z (t) = (\Z_0 (t))^N \ ,
\label{eq:z0n}
\end{equation}
where
\begin{equation}
    \Z_0 (t) = \sum_{\{ \alpha_n \}} \mel{\alpha_n}{T \exp \left( - i \int_0^t d \tau' \hat{H}_n (\tau') \right)}{\alpha_n} \ .
\label{eq:z0}
\end{equation}
Here $\{ \ket{\alpha_n} \}$ is a set of basis states for the fermion mode indexed by $n$. One choice of basis here is the BCS basis, for which $\{ \ket{\alpha_n} \} = \{ \ket{11}_n , \ket{00}_n \}$. Here the two basis states correspond respectively to the presence or absence of a Cooper pair in the $n-$fermionic mode. $\Z_0 (t)$ is the same for all $n$ since in this basis, the elements of the Hamiltonian matrix $\hat{H}_n (\tau)$ \eqref{eq:hntau} are independent of $n$.

The saddle point equation for $\Delta (\tau)$ is obtained by taking the functional derivative of the action in \eqref{eq:action} with respect to $\bar{\Delta}(\tau)$ and setting it to zero. This yields the saddle point,
\begin{equation}
    \Delta (\tau) = g \frac{\Z_1 (\tau; t)}{\Z_0 (t)} \ ,
\label{eq:del}
\end{equation}
where
\begin{equation}
    \Z_1 (\tau; t) = \sum_{\{ \alpha_n \}} \mel{\alpha_n}{\left(T \exp \left( - i \int_\tau^t d \tau' \hat{H}_n (\tau') \right) \right) \ha_{n \dn} \ha_{n \up} 
    \left(T \exp \left( - i \int_0^{\tau} d \tau' \hat{H}_n (\tau') \right) \right)}{\alpha_n} \ ,
\label{eq:z1}
\end{equation}
and $\Z_0 (t)$ is given by \eqref{eq:z0}.
In a similar way, we obtain the saddle point equation for $\bar{\Delta}(\tau)$ as 
\begin{equation}
    \bar{\Delta} (\tau) = g \frac{\bar{\Z}_1 (\tau; t)}{\Z_0 (t)} \ ,
\label{eq:delbar}
\end{equation}
where
\begin{equation}
    \bar{\Z}_1 (\tau; t) = \sum_{\{ \alpha_n \}} \mel{\alpha_n}{\left(T \exp \left( - i \int_\tau^t d \tau' \hat{H}_n (\tau') \right) \right) \ha_{n \up}\dg \ha_{n \dn}\dg 
    \left(T \exp \left( - i \int_0^{\tau} d \tau' \hat{H}_n (\tau') \right) \right)}{\alpha_n} \ .
\label{eq:z1bar}
\end{equation}
We see that $\Delta (\tau)$ and $\bar{\Delta} (\tau)$ are not complex conjugates in general. These saddle points have explicit dependence on $\tau$. Hence, the mean-field Hamiltonian $\hat{H}_n$ defined in \eqref{eq:hntau} is non-Hermitian and time($\tau$)-dependent.

\subsection{Time dependence of mean fields}
The standard way to solve these saddle point equations would be an iterative procedure. We would begin by choosing arbitrary fields $\Delta(\tau)$ and $\bar{\Delta}(\tau)$ for all $\tau \in (0, t)$. We would calculate $\Z_0 (t)$, $\Z_1 (\tau; t)$ and $\bar{\Z}_1 (\tau; t)$ with our chosen fields $\Delta (\tau)$ and $\bar{\Delta} (\tau)$. Then we use these values of $\Z_0 (t)$, $\Z_1 (\tau; t)$ and $\bar{\Z}_1 (\tau; t)$ to calculate new fields $\Delta (\tau)$ and $\bar{\Delta} (\tau)$ according to \eqref{eq:del} and \eqref{eq:delbar}. We use these new fields $\Delta (\tau)$ and $\bar{\Delta} (\tau)$ to again calculate $\Z_0$, $\Z_1$ and $\bar{\Z}_1$. We get a self-consistent solution for the saddle point equations if the sequence of fields $(\Delta (\tau), \bar{\Delta} (\tau))$ thus generated, converges.

This standard procedure is computationally cumbersome. Also it is not certain that such an iteration would converge. For example, if $\Z_0 (t)$ becomes sufficiently close to $0$ at some step of the iteration, then the values for $\Delta (\tau)$ and $\bar{\Delta} (\tau)$ diverge, and the iterative procedure breaks down.

However, it turns out that we can do better for this flat-band model. We take the derivative of $\Delta(\tau)$ with $\tau$. From \eqref{eq:del},
\begin{equation}
    \frac{\d \Delta (\tau)}{\d \tau} 
    = \frac{g}{\Z_0 (t)} \lim_{h \rar 0} \frac{\Z_1(\tau + h; t) - \Z_1 (\tau; t)}{h} \ .
\end{equation}
Using the expression for $\Z_1 (\tau ; t)$ from \eqref{eq:z1}, this becomes
\begin{equation}
\begin{split}
    \frac{\d \Delta (\tau)}{\d \tau} = & \frac{g}{\Z_0 (t)} \lim_{h \rar 0} \frac{1}{h} \sum_{\{ \a_n \}} \bra{\a_n} \left(T \exp \left( - i \int_{\tau + h}^t d \tau' \hat{H}_n (\tau') \right) \right) \ha_{n \dn} \ha_{n \up} 
    \left(T \exp \left( - i \int_0^{\tau + h} d \tau' \hat{H}_n (\tau') \right) \right) \\
    & - \left(T \exp \left( - i \int_\tau^t d \tau' \hat{H}_n (\tau') \right) \right) \ha_{n \dn} \ha_{n \up} 
    \left(T \exp \left( - i \int_0^{\tau} d \tau' \hat{H}_n (\tau') \right) \right) \ket{ \a_n} \ .
\end{split}
\end{equation}
This evaluates to
\begin{equation}
\begin{split}
    \frac{\d \Delta(\tau)}{\d \tau} = & \frac{ i g}{\Z_0 (t)} \sum_{\{ \a_n \}} \bra{\a_n} 
    \left(T \exp \left( - i \int_{\tau}^t d \tau' \hat{H}_n (\tau') \right) \right) \\
    &[ \hat{H}_n (\tau), \ha_{n \dn} \ha_{n \up} ]
    \left(T \exp \left( - i \int_0^{\tau} d \tau' \hat{H}_n (\tau') \right) \right) \ket{\a_n} \ .
\end{split}
\label{eq:ddel1}
\end{equation}
Referring to the mean-field Hamiltonian in \eqref{eq:hntau}, we find that the commutator appearing above evaluates to
\begin{equation}
    [ \hat{H}_n (\tau), \ha_{n \dn} \ha_{n \up} ] = 
    - 2 \epsilon  \ha_{n \dn} \ha_{n \up} - \Delta (\tau) (\ha_{n \up}\dg \ha_{n \up} + \ha_{n \dn}\dg \ha_{n \dn} - 1) \ .
\label{eq:comdel}
\end{equation}

Now let us define the generalized expectation value of the Cooper pair `number' operator (analogous to $\Delta (\tau)$ in \eqref{eq:del}) as
\begin{equation}
    n(\tau) = \frac{\Z_2 (\tau ; t)}{\Z_0 (t)} \ ,
\label{eq:n}
\end{equation}
where $\Z_2 (\tau; t)$ is defined as
\begin{equation}
\begin{split}
    \Z_2 (\tau; t) = &\sum_{\{ \alpha_n \}} \bra{\alpha_n}\left(T \exp \left( - i \int_\tau^t d \tau' \hat{H}_n (\tau') \right) \right) \cdot \\
    &(\ha_{n \up}\dg \ha_{n \up} + \ha_{n \dn}\dg \ha_{n \dn} - 1)
    \left(T \exp \left( - i \int_0^{\tau} d \tau' \hat{H}_n (\tau') \right) \right) \ket{\alpha_n} \ ,
\end{split}
\label{eq:z2}
\end{equation}
and $\Z_0 (t)$ is given by \eqref{eq:z0}. Using this definition of $n(\tau)$ and the commutator \eqref{eq:comdel}, \eqref{eq:ddel1} reduces to 
\begin{equation}
    \frac{\d \Delta(\tau)}{\d \tau} = - i \Delta (\tau) (2 \epsilon + g \, n(\tau)) \ .
\label{eq:ddel2}
\end{equation}
Thus we see that in the self-consistent mean field formalism, the time($\tau$)-evolution equation for $\Delta (\tau)$ depends only on the concurrent values of the fields $\Delta (\tau)$ and $n (\tau)$.

Following the same procedure, we get the time-evolution equation for $\bar{\Delta} (\tau)$ as 
\begin{equation}
    \frac{\d \bar{\Delta}(\tau)}{\d \tau} =  i \bar{\Delta} (\tau) (2 \epsilon + g \, n(\tau)) \ .
\label{eq:ddelbar2}
\end{equation}

Finally, we can find the time-evolution equation for $n(\tau)$ by differentiating $\Z_2 (\tau; t)$ with $\tau$ from \eqref{eq:z2}. We get
\begin{equation}
\begin{split}
    \frac{\d n(\tau)}{\d \tau} = & \frac{ i}{\Z_0 (t)} \sum_{\{ \a_n \}} \bra{\a_n} 
    \left(T \exp \left( - i \int_{\tau}^t d \tau' \hat{H}_n (\tau') \right) \right) \\
    &[ \hat{H}_n (\tau), \ha_{n \up}\dg \ha_{n \up} + \ha_{n \dn}\dg \ha_{n \dn} - 1 ]
    \left(T \exp \left( - i \int_0^{\tau} d \tau' \hat{H}_n (\tau') \right) \right) \ket{\a_n} \ .
\end{split}
\label{eq:dn1}
\end{equation}
Using \eqref{eq:hntau} the commutator above yields
\begin{equation}
    [ \hat{H}_n (\tau), \ha_{n \up}\dg \ha_{n \up} + \ha_{n \dn}\dg \ha_{n \dn} - 1 ] = 2 \Delta (\tau) \ha_{n \up}\dg \ha_{n \dn}\dg - 2 \bar{\Delta} (\tau) \ha_{n \dn} \ha_{n \up} \ .
\end{equation}
Substituting this into \eqref{eq:dn1}, we get
\begin{equation}
    \frac{\d n(\tau)}{\d \tau} = \frac{2 i }{g} (\Delta (\tau) \bar{\Delta} (\tau) - \bar{\Delta}(\tau) \Delta (\tau)) = 0 \ .
\end{equation}
We find that $n (\tau)$ is constant as a function of $\tau$. Let us call this constant $n$.
\begin{equation}
    n = n (\tau) = n (0) = \frac{\Z_2 (0; t)}{\Z_0 (t)} \quad \text{[from \eqref{eq:n}]}.
\label{eq:n0}
\end{equation}
Using this in \eqref{eq:ddel2} and \eqref{eq:ddelbar2}, we obtain
\begin{equation}
    \Delta (\tau) = \Delta_0 e^{-i(2 \epsilon + g \, n) \tau} \quad ,
    \quad \bar{\Delta}(\tau) = \bar{\Delta}_0 e^{i(2 \epsilon + g \, n) \tau} \ ,
\label{eq:deltau}
\end{equation}
where $\Delta_0 = \Delta(\tau = 0)$ and $\bar{\Delta}_0 = \bar{\Delta}(\tau = 0)$ are the initial values of the fields. Thus we find that in this simple flat-band model, the time($\tau$)-dependence of the mean fields can be expressed exactly.

\subsection{Self-consistent solution of saddle point equations}

To calculate the spectral form factor, we need to determine the values of $n$, $\Delta_0$ and $\bar{\Delta}_0$ self-consistently. Using \eqref{eq:deltau}, the mean-field Hamiltonian \eqref{eq:hntau} can be written in the BCS basis as
\begin{equation}
    \hat{H}_n (\tau) = 
    \begin{pmatrix}[1.0]
        \epsilon & - \Delta_0 e^{-i(2 \epsilon + g \, n) \tau} \\
        - \bar{\Delta}_0 e^{i(2 \epsilon + g \, n) \tau} & - \epsilon
    \end{pmatrix}
    + \frac{\bar{\Delta}_0 \Delta_0}{g} \mathbf{1} \ .
\end{equation}
The time($\tau$)-dependence of this Hamiltonian can be factored by going into a `rotating' basis,
\begin{equation}
    \hat{H}_n (\tau) = \hat{U}_n^{-1} (\tau) \hat{H}_n \hat{U}_n (\tau) \ ,
\label{eq:udef}
\end{equation}
where
\begin{equation}
    \hat{H}_n =
    \begin{pmatrix}[1.0]
        \epsilon & -\Delta_0 \\
        - \bar{\Delta}_0 & - \epsilon 
    \end{pmatrix}
    + \frac{\bar{\Delta}_0 \Delta_0}{g} \mathbf{1} \quad ,\quad 
    \hat{U}_n (\tau) = 
    \begin{pmatrix}[1.0]
        1 & 0 \\
        0 & e^{-i(2 \epsilon + g \, n) \tau}
    \end{pmatrix}.
\label{eq:uval}
\end{equation}
Note that $\hat{U}_n (\tau)$ is not in general a unitary matrix, since $n$ as defined in \eqref{eq:n0} is not real-valued.

The time-evolution operator can be written as
\begin{equation}
T \exp \left( - i \int_0^t d \tau' \hat{H}_n (\tau') \right)
= \lim_{\delta \tau' \rar 0} T \prod_{m = 1}^{t/\delta \tau'} \exp(-i \hat{H}_n (m \delta \tau') \delta \tau') \ .
\end{equation}
Using \eqref{eq:udef} this becomes
\begin{equation}
    T \exp \left( - i \int_0^t d \tau' \hat{H}_n (\tau') \right)
    = \lim_{\delta \tau' \rar 0} T \prod_{m = 1}^{t/\delta \tau'}
    \exp (-i \delta \tau' \hat{U}_n^{-1} (m \delta \tau') \hat{H}_n \hat{U}_n (m \delta \tau') ) \ ,
\end{equation}
giving
\begin{equation}
    T \exp \left( - i \int_0^t d \tau' \hat{H}_n (\tau') \right)
    = \lim_{\delta \tau' \rar 0} T \prod_{m = 1}^{t/\delta \tau'}
    \hat{U}_n^{-1} (m \delta \tau') \exp(-i \hat{H}_n \delta \tau') \hat{U}_n (m \delta \tau') \ .
\label{eq:ht1}
\end{equation}
To evaluate this product we note that
\begin{equation}
    \hat{U}_n ((m+1) \delta \tau')\, \hat{U}_n^{-1} (m \delta \tau') = 
    \begin{pmatrix}[1.0]
        1 & 0 \\
        0 & e^{- i (2 \epsilon + g\, n) \delta \tau'}
    \end{pmatrix}
    = \hat{U}_n (\delta \tau') \ .
\end{equation}
\eqref{eq:ht1} reduces to 
\begin{equation}
    T \exp \left( - i \int_0^t d \tau' \hat{H}_n (\tau') \right)
    = \lim_{\delta \tau' \rar 0} \hat{U}_n^{-1} (t) e^{-i \hat{H}_n \delta \tau'} \left( \hat{U}_n (\delta \tau') e^{-i \hat{H}_n \delta \tau'} \right)^{t/\delta \tau' - 1} \hat{U}_n (0) \ .
\end{equation}
We can write
\begin{equation}
    \hat{U}_n (\delta \tau') e^{-i \hat{H}_n \delta \tau'} = \mathbf{1} + i A \delta \tau' \ ,
\end{equation}
where
\begin{equation}
    A = \begin{pmatrix}[1.0]
        - \epsilon &  \Delta_0 \\
         \bar{\Delta}_0 & - (\epsilon + g\, n)
    \end{pmatrix}
     - \frac{\bar{\Delta}_0 \Delta_0}{g} \mathbf{1},
\end{equation}
so that the time evolution operator becomes
\begin{equation}
     T \exp \left( - i \int_0^t d \tau' \hat{H}_n (\tau') \right)
     = \hat{U}_n^{-1} (t) \exp (i A t) \hat{U}_n (0) \ .
\label{eq:hta}
\end{equation}
To calculate the exponent of $A$, we diagonalize it. The eigenvalues of $A$ are
\begin{equation}
\begin{split}
    \lambda_{\pm} &= -  \left( \epsilon + \frac{g\, n}{2} + \frac{\bar{\Delta}_0 \Delta_0}{g} \right) \pm  D \quad ,\quad \text{where} \\
    D &= \sqrt{\bar{\Delta}_0 \Delta_0 + \frac{g^2\, n^2}{4}} \ .
\end{split}
\label{eq:eval}
\end{equation}
We work with the case $ D \neq 0$ in which case the eigenvalues are distinct and the matrix $A$ is diagonalizable. The case $D = 0$ can be evaluated in a separate calculation not presented here; doing that calculation, we see that the result for $D = 0$ coincides with that for $D \neq 0$ in the limit $D \rar 0$. Hence let us proceed with our calculation for $D \neq 0$. The matrix of eigenvectors corresponding to the eigenvalues $\lambda_+$ and $\lambda_-$ is
\begin{equation}
    V = \begin{pmatrix}[1.0]
        \frac{g\, n}{2} + D & - \Delta_0 \\
        \bar{\Delta}_0 & \frac{g n}{2} + D
    \end{pmatrix} .
\end{equation}
The exponential of $A$ can then be written as
\begin{equation}
    \exp (i A \tau) = V \begin{pmatrix}[1.0]
        e^{i \lambda_+ \tau} & 0 \\
        0 & e^{i \lambda_- \tau}
    \end{pmatrix} V^{-1} \ .
\label{eq:ea}
\end{equation}
Substituting this in \eqref{eq:hta} and using \eqref{eq:eval}, we get
\begin{equation}
\begin{split}
& T \exp \left( - i \int_0^t d \tau' \hat{H}_n (\tau') \right) =   e^{- i \frac{\bar{\Delta}_0 \Delta_0}{g} t }\, \cdot \\
& \begin{pmatrix}
    e^{- i \left( \epsilon + \frac{g\, n}{2} \right) t}
    \left( \cos (D t) + i \frac{g\, n}{2 D} \sin (D t) \right) &
    e^{- i \left( \epsilon + \frac{g\, n}{2} \right) t} \,
    \frac{i \Delta_0}{D} \sin (D t) \\
    e^{ i \left( \epsilon + \frac{g\, n}{2} \right) t} \,
    \frac{i \bar{\Delta}_0}{D} \sin (D t) &
    e^{ i \left( \epsilon + \frac{g\, n}{2} \right) t}
    \left( \cos (D t) - i \frac{g\, n}{2 D} \sin (D t) \right)
\end{pmatrix} .
\end{split}
\label{eq:htf}
\end{equation}
The spectral form factor (for a single fermionic mode) is just the trace of this time evolution operator as in \eqref{eq:z0}. Thus we get
\begin{equation}
    \Z_0 (t) = e^{- i \frac{\bar{\Delta}_0 \Delta_0}{g} t } \left[ 2 \cos (D t) \cos \left( \left( \epsilon + \frac{g \, n}{2} \right) t \right)
 + \frac{g \, n}{D} \sin (D t) \sin \left( \left( \epsilon + \frac{g \, n}{2} \right) t \right) \right]\ .
\label{eq:z0f}
\end{equation}

Doing similar calculations as those from \eqref{eq:hta}-\eqref{eq:htf}, we get the remaining time-evolution operators as 
\begin{equation}
\begin{split}
    & T \exp \left( - i \int_0^{\tau} d \tau' \hat{H}_n (\tau') \right)
    = \hat{U}_n^{-1} (\tau) \exp (i A \tau) \hat{U}_n (0)
    = e^{- i \frac{\bar{\Delta}_0 \Delta_0}{g} \tau }\, \cdot \\
    & \begin{pmatrix}
    e^{- i \left( \epsilon + \frac{g\, n}{2} \right) \tau}
    \left( \cos (D \tau) + i \frac{g\, n}{2 D} \sin (D \tau) \right) &
    e^{- i \left( \epsilon + \frac{g\, n}{2} \right) \tau} \,
    \frac{i \Delta_0}{D} \sin (D \tau) \\
    e^{ i \left( \epsilon + \frac{g\, n}{2} \right) \tau} \,
    \frac{i \bar{\Delta}_0}{D} \sin (D \tau) &
    e^{ i \left( \epsilon + \frac{g\, n}{2} \right) \tau}
    \left( \cos (D \tau) - i \frac{g\, n}{2 D} \sin (D \tau) \right)
\end{pmatrix} \ ,
\end{split}
\label{eq:htauf}
\end{equation}
and
\begin{equation}
\begin{split}
      & T \exp \left( - i \int_{\tau}^t d \tau' \hat{H}_n (\tau') \right)
     = \hat{U}_n^{-1} (t) \exp (i A (t - \tau)) \hat{U}_n (\tau)
     = e^{- i \frac{\bar{\Delta}_0 \Delta_0}{g} (t - \tau)}\, \cdot \\
     & \begingroup % keep the change local
\setlength\arraycolsep{-7pt} \begin{pmatrix}
    e^{- i \left( \epsilon + \frac{g\, n}{2} \right) (t-\tau)}
    \left( \cos (D (t-\tau)) + i \frac{g\, n}{2 D} \sin (D (t-\tau)) \right) &
    e^{- i \left( \epsilon + \frac{g\, n}{2} \right) (t+\tau)} \,
    \frac{i \Delta_0}{D} \sin (D (t-\tau)) \\
    e^{ i \left( \epsilon + \frac{g\, n}{2} \right) (t+\tau)} \,
    \frac{i \bar{\Delta}_0}{D} \sin (D (t-\tau)) &
    e^{ i \left( \epsilon + \frac{g\, n}{2} \right) (t-\tau)}
    \left( \cos (D (t-\tau)) - i \frac{g\, n}{2 D} \sin (D (t-\tau)) \right)
\end{pmatrix} \endgroup
\end{split}
\label{eq:htautf}
\end{equation}
Substituting these time-evolution operators in \eqref{eq:z1}, we get
\begin{equation}
\begin{split}
    & \Z_1 (\tau; t) =  e^{- i \frac{\bar{\Delta}_0 \Delta_0}{g} t }
    e^{-i (2 \epsilon + g\, n) \tau} \, \frac{i \Delta_0}{D} \bigg[ \cos \left( \left( \epsilon + \frac{g\, n}{2} \right) t \right) \sin (D t) \\
    &+ i \sin \left( \left( \epsilon + \frac{g\, n}{2} \right) t \right)
    \sin (D (2 \tau - t))
    + \frac{g\, n}{D} \sin (D \tau) \sin (D (t - \tau)) \sin \left( \left( \epsilon + \frac{g\, n}{2} \right) t \right)
    \bigg]\ .
\end{split}
\label{eq:z1f}
\end{equation}
Also, \eqref{eq:z2} gives
\begin{equation}
\begin{split}
    \Z_2 (\tau; t) = e^{- i \frac{\bar{\Delta}_0 \Delta_0}{g} t }
    \left[ -2 i \cos (D t) \sin \left( \left( \epsilon + \frac{g\, n}{2} \right) t \right)
    + \frac{i g\, n}{D} \sin (D t) \cos \left( \left( \epsilon + \frac{g\, n}{2} \right) t \right) \right]\ .
\end{split}
\label{eq:z2f}
\end{equation}

To get the self-consistency condition for the mean-field we substitute the expressions for $\Z_0 (t)$, $\Z_1 (\tau; t)$ from \eqref{eq:z0f} and \eqref{eq:z1f} into \eqref{eq:del}, giving
\begin{equation}
\begin{split}
    & \left[\cos \left( \left( \epsilon + \frac{g\, n}{2} \right) t \right) \left( 2 \cos (D t) - \frac{i g}{D} \sin (D t) \right)
    + \sin \left( \left( \epsilon + \frac{g\, n}{2} \right) t \right) \frac{g\, n}{D} \left( \sin (D t) + \frac{i g}{D} \cos (D t) \right) \right] \Delta_0 \\
    & = -\frac{g}{D} \sin \left( \left( \epsilon + \frac{g\, n}{2} \right) t \right)
    \left[ \sin (D (2 \tau - t)) - i \frac{g\, n}{D} \cos (D (2 \tau - t)) \right] \Delta_0 \ .
\end{split}
\label{eq:sced}
\end{equation}
We notice that the left hand side of the above equation is independent of the intermediate time $\tau$, whereas the right hand side explicitly depends on it. Thus, we must have 
\begin{equation}
\Delta_0 = 0 \quad \text{or} \quad \sin \left( \left( \epsilon + \frac{g\, n}{2} \right) t \right) = 0 \ .
\label{eq:scd}
\end{equation}
It is also possible to have $D = 0$. But then the eigenvalues in \eqref{eq:eval} are not distinct and the matrix $A$ cannot be diagonalized. As mentioned after \eqref{eq:eval}, we looked at this case separately and found that it leads to the same conditions as above.

In \eqref{eq:z1f} we calculated $\Z_1 (\tau; t)$ and used it in the self-consistency equation for $\Delta (\tau)$. In a similar manner we calculate $\bar{\Z}_1 (\tau; t)$ and use it to get the self-consistency condition for $\bar{\Delta} (\tau)$. It turns out that analogous to \eqref{eq:scd}, this gives
\begin{equation}
    \bar{\Delta}_0 = 0 \quad \text{or} \quad \sin \left( \left( \epsilon + \frac{g\, n}{2} \right) t \right) = 0 \ .
\end{equation}

The final self-consistency condition is that for $n(\tau)$ in \eqref{eq:n}. Substituting from \eqref{eq:z0f} and \eqref{eq:z2f}, we get
\begin{equation}
\begin{split}
    & 2 \cos (D t) \left[ n \cos \left( \left( \epsilon + \frac{g\, n}{2} \right) t \right) + i \sin \left( \left( \epsilon + \frac{g\, n}{2} \right) t \right) \right] \\
    & =  \frac{g\, n}{D} \sin (D t) \left[ i \cos \left( \left( \epsilon + \frac{g\, n}{2} \right) t \right) - n \sin \left( \left( \epsilon + \frac{g\, n}{2} \right) t \right) \right] \ .
\end{split}
\label{eq:scn}
\end{equation}
We want to see if \eqref{eq:scd} can satisfy this self-consistency condition for $n(\tau)$. If we have $\Delta_0 = 0$ then from \eqref{eq:eval}, we have $D = \pm g\, n/2$. \eqref{eq:scn} becomes
\begin{equation}
\begin{split}
    & 2 n \left[ \cos \left( \left( \epsilon + \frac{g\, n}{2} \right) t \right) \cos \left( \frac{g\, n}{2} t \right) + 
    \sin \left( \left( \epsilon + \frac{g\, n}{2} \right) t \right) \sin \left( \frac{g\, n}{2} t \right) \right] \\
    & =  2 i \left[ \cos \left( \left( \epsilon + \frac{g\, n}{2} \right) t \right) \sin \left( \frac{g\, n}{2} t \right) - 
    \sin \left( \left( \epsilon + \frac{g\, n}{2} \right) t \right) \cos \left( \frac{g\, n}{2} t \right) \right] \ .
\end{split}
\end{equation}
This reduces to 
\begin{equation}
    n = - i \tan (\epsilon t) \ .
\end{equation}
Thus we have
\begin{equation}
    \Delta_0 = \bar{\Delta}_0 = 0 \ ,\ n = - i \tan (\epsilon t)
\label{eq:sp1}
\end{equation}
as a self-consistent saddle point.

With the other solution in  \eqref{eq:scd}, $\sin ((\epsilon + g\, n/2) t) = 0$, the self-consistency equation \eqref{eq:sced} for $\Delta$ and equation \eqref{eq:scn} for $n$, both reduce to
\begin{equation}
    \frac{i}{2} \frac{\tan (D t)}{D} = \frac{1}{g} \ .
\label{eq:spd}
\end{equation}
Substituting for $D$ from \eqref{eq:eval}, we find that
\begin{equation}
n = \frac{2}{g} \left( \frac{k \pi}{t} - \epsilon \right) \ ,\ 
\frac{i}{2 \sqrt{\bar{\Delta}_0 \Delta_0 + \left( \frac{k \pi}{t} - \epsilon \right)^2} } \tan \left( t \sqrt{\bar{\Delta}_0 \Delta_0 + \left( \frac{k \pi}{t} - \epsilon \right)^2} \right) = \frac{1}{g}
\label{eq:sp2}
\end{equation}
is a valid saddle point for any integer $k$. The time-evolution equation for $\Delta$ and $\bar{\Delta}$, \eqref{eq:deltau}, becomes
\begin{equation} \label{eq:timedep}
    \Delta (\tau) = \Delta_0 \exp \left( - i \frac{2 \pi k \tau}{t} \right) \ ,\ 
    \bar{\Delta} (\tau) = \bar{\Delta}_0 \exp \left(  i \frac{2 \pi k \tau}{t} \right) \ .
\end{equation}
We see that the self-consistency equation enforces $\Delta (t) = \Delta (0)$ and $\bar{\Delta} (t) = \bar{\Delta} (0)$. The choice $k = 0$ gives a time($\tau$)-independent saddle point.

\end{document}